\documentclass[]{aastex63}

\shorttitle{MB06074: xallarap contaminants}
\shortauthors{Rota et al.}

\begin{document}

\title{MOA-2006-BLG-074: recognizing xallarap contaminants in planetary microlensing}

\correspondingauthor{Valerio Bozza}
\email{valboz@sa.infn.it}

\author{P. Rota}
\affiliation{Dipartimento di Fisica E.R. Caianiello, Universit\`a di Salerno, Via Giovanni Paolo II 132, 84084, Fisciano, Italy}
\affiliation{Istituto Nazionale di Fisica Nucleare, Sezione di Napoli, 80126, Napoli, Italy}

\author{Y. Hirao}
\affiliation{Department of Earth and Space Science, Graduate School of Science, Osaka University, Toyonaka, Osaka 560-0043, Japan}

\author{V. Bozza}
\affiliation{Dipartimento di Fisica E.R. Caianiello, Universit\`a di Salerno, Via Giovanni Paolo II 132, 84084, Fisciano, Italy}
\affiliation{Istituto Nazionale di Fisica Nucleare, Sezione di Napoli, 80126, Napoli, Italy}

\author{F. Abe}
\affiliation{Institute for Space-Earth Environmental Research, Nagoya University, Nagoya 464-8601, Japan}
\author{R. Barry}
\affiliation{Code 667, NASA Goddard Space Flight Center, Greenbelt, MD 20771, USA}
\author{David P.~Bennett}
\affiliation{Code 667, NASA Goddard Space Flight Center, Greenbelt, MD 20771, USA}
\affiliation{Department of Astronomy, University of Maryland, College Park, MD 20742, USA}
\author{A. Bhattacharya}
\affiliation{Code 667, NASA Goddard Space Flight Center, Greenbelt, MD 20771, USA}
\affiliation{Department of Astronomy, University of Maryland, College Park, MD 20742, USA}
\author{Ian A. Bond}
\affiliation{Institute of Natural and Mathematical Sciences, Massey University, Auckland 0745, New Zealand}
\author{M. Donachie}
\affiliation{Department of Physics, University of Auckland, Private Bag 92019, Auckland, New Zealand}
\author{A. Fukui}
\affiliation{Department of Earth and Planetary Science, Graduate School of Science, The University of Tokyo, 7-3-1 Hongo, Bunkyo-ku, Tokyo 113-0033, Japan}
\affiliation{Instituto de Astrof\'isica de Canarias, V\'ia L\'actea s/n, E-38205 La Laguna, Tenerife, Spain}
\author{H. Fujii}
\affiliation{Institute for Space-Earth Environmental Research, Nagoya University, Nagoya 464-8601, Japan}

\author{S. Ishitani Silva}
\affiliation{Department of Physics, The Catholic University of America, Washington, DC 20064, USA}
\affiliation{Code 667, NASA Goddard Space Flight Center, Greenbelt, MD 20771, USA}
\author{Y. Itow}
\affiliation{Institute for Space-Earth Environmental Research, Nagoya University, Nagoya 464-8601, Japan}
\author{R. Kirikawa}
\affiliation{Department of Earth and Space Science, Graduate School of Science, Osaka University, Toyonaka, Osaka 560-0043, Japan}
\author{Naoki Koshimoto}
\affiliation{Department of Astronomy, Graduate School of Science, The University of Tokyo, 7-3-1 Hongo, Bunkyo-ku, Tokyo 113-0033, Japan}
\author{M. C. A. Li}
\affiliation{Department of Physics, University of Auckland, Private Bag 92019, Auckland, New Zealand}
\author{Y. Matsubara}
\affiliation{Institute for Space-Earth Environmental Research, Nagoya University, Nagoya 464-8601, Japan}
\author{S. Miyazaki}
\affiliation{Department of Earth and Space Science, Graduate School of Science, Osaka University, Toyonaka, Osaka 560-0043, Japan}
\author{Y. Muraki}
\affiliation{Institute for Space-Earth Environmental Research, Nagoya University, Nagoya 464-8601, Japan}
\author{G. Olmschenk}
\affiliation{Code 667, NASA Goddard Space Flight Center, Greenbelt, MD 20771, USA}
\author{C. Ranc}
\affiliation{Sorbonne Universit\'e, CNRS, UMR 7095, Institut d'Astrophysique de Paris, 98 bis bd Arago, 75014 Paris, France}
\author{Y. Satoh}
\affiliation{Department of Earth and Space Science, Graduate School of Science, Osaka University, Toyonaka, Osaka 560-0043, Japan}
\author{Takahiro Sumi}
\affiliation{MOA collaboration}
\affiliation{Department of Earth and Space Science, Graduate School of Science, Osaka University, Toyonaka, Osaka 560-0043, Japan}
\author{D. Suzuki}
\affiliation{Department of Earth and Space Science, Graduate School of Science, Osaka University, Toyonaka, Osaka 560-0043, Japan}
\author{P. J. Tristram}
\affiliation{University of Canterbury Mt.\ John Observatory, P.O. Box 56, Lake Tekapo 8770, New Zealand}
\author{A. Yonehara}
\affiliation{Department of Physics, Faculty of Science, Kyoto Sangyo University, 603-8555 Kyoto, Japan}

\begin{abstract}

MOA-2006-BLG-074 was selected as one of the most promising planetary candidates in a retrospective analysis of the MOA collaboration: its asymmetric high-magnification peak can be perfectly explained by a source passing across a central caustic deformed by a small planet. However, after a detailed analysis of the residuals, we have realized that a single lens and a source orbiting with a faint companion provides a more satisfactory explanation for all the observed deviations from a Paczynski curve and the only physically acceptable interpretation. Indeed the orbital motion of the source is constrained enough to allow a very good characterization of the binary source from the microlensing light curve. The case of MOA-2006-BLG-074 suggests that the so-called xallarap effect must be taken seriously in any attempts to obtain accurate planetary demographics from microlensing surveys.
\end{abstract}


\section{Introduction} \label{sec:intro}

Gravitational microlensing is a powerful technique to discover planets that are hidden to other methods such as transits or radial velocity \citep{MaoPaczynski1991,GouldLoeb1992,BennettRhie1996}. Thanks to microlensing we can detect planets in intermediate-wide orbits, allowing the exploration of planetary systems in regions beyond the snow line, which would remain barely touched otherwise. For this reason, the hundred of planets discovered by this method is particularly precious for the construction of a broad statistics of planets that are formed by accretion of ice particles.

A microlensing event appears as a symmetric bell-shaped peak in the light curve of a distant source with a duration ranging from a few hours to hundreds of days \citep{Mroz2017}. Planets show up with a variety of anomalies, ranging from short additional peaks or dips to longer distortions, depending on the geometry of the caustics. It is clear that the most common case of short duration additional peaks could be confused with the case of a secondary source suffering microlensing \citep{GriestHu1992}. A source consisting in a binary system should thus lead to the sum of two Paczynski curves, if both sources are luminous enough. Nevertheless, the orbital motion of the two sources induces additional distortions to the light curve, with periodic modulations induced by the oscillatory motion of the two stars around the common center of mass \citep{Paczynski1997}. The periods of binary systems may range from a few hours for very close systems to hundreds of years for widely spaced pairs. Furthermore, given the steepness of the luminosity-mass relation, in most cases the secondary star might be much fainter than the primary, while still being able to perturb its trajectory. For this reason, the binarity of the source may sometimes arise just from the perturbation of the motion of the primary without any signs of additional light. Early examples of possible events observed by the EROS and MACHO projects with modulations induced by the orbital motion of the source around a hidden companion are in \citet{Palanque1998,Derue1999,Alcock2001}.
Such effect is very similar to annual parallax \citep{Gould1992,Alcock1995}, arising from the orbital motion of the Earth around the Sun, save for the fact that it involves the other end of the line of sight. For this reason, it is commonly referred as xallarap (i.e. the inverse of parallax). It was soon realized that the two effects may produce identical long-term perturbations if the duration of the event is short compared to the orbital period \citep{Smith2003}. Indeed, for many microlensing events of interest, models with xallarap compete with models with parallax and cannot be excluded on the basis of physical arguments \citep{Ghosh2004,Bennett2008,Hwang2010,Miyake2012,Furusawa2013,Koshimoto2017}. However, the combination of constraints from the third Kepler's law with other information on the source may sometimes pinpoint the masses of the two components \citep{HanGould1997}. This may help rule out the xallarap model if the companion has a mass incompatible with the observed flux \citep{Han2013,Kains2013,Kim2021}, unless it is a black hole. Furthermore, if a xallarap model returns a period of 1 year, there is a high chance that the fit has converged to a mirror solution of a parallax model \citep{Dong2009,Hwang2011}.

Nevertheless, binary stars are common around the Galaxy, with a multiplicity fraction around $40\%$ \citep{Lada2006,Badenes2018}, which should be probably further increased for the stars in the bulge targeted by microlensing campaigns. Although most of these binaries are too wide to show any signatures in microlensing events, a systematic study of 22 long-duration events in the bulge showed that $23\%$ of them were actually affected by xallarap \citep{Poindexter2005}. Indeed, some planetary microlensing events require binary sources for a complete modeling of all deviations \citep{Sumi2010,Bennett2018,Miyazaki2020}. The impact of xallarap in the planet detection efficiencies of microlensing surveys has been poorly investigated, but it is known to exist \citep{Zhu2017}.

The study of binary sources and their orbital motion may even provide new opportunities: it may help break degeneracies in microlensing \citep{HanGould1997}, or it may open a new channel to discover planets orbiting around the source star \citep{RahvarDominik2009,Bagheri2019,Miyazaki2021}. Indeed, xallarap must be considered as a possible alternative before introducing additional bodies to explain the observed deviations \citep{Hwang2018}.

In the following paper, we present the analysis of the planetary microlensing event MOA-2006-BLG-074, whose anomaly has been initially overlooked and only recently noticed in a retrospective analysis by the MOA collaboration. This event provides an enlightening example of a very promising planetary candidate that can be actually explained by the xallarap effect. Understanding events like this is important to develop methods to discriminate between the two cases in future automatic analysis pipelines running on large datasets. 

In section 2 we present the observations from MOA telescope. Section 3 explains the modeling stages in detail. In section 4 we show the analysis of the source star in which we derive the angular Einstein radius. Then, in section 5 we use all available information to characterize both the lens and the source system. Finally, in section 6 we draw our conclusions.

\section{Observations} \label{sec:obs}

The event was detected by the Microlensing Observations in Astrophysics (MOA; \citep{MOA2001,MOA2003}) at the J2000 equatorial coordinates (RA,Dec) = $(18^h 05^m 27^s.341, -31^{\degr} 47' 17''.68)$ corresponding to Galactic coordinates ($\textit{l,b}$) = $(-0.316 \degr, -5.131\degr)$. Several years of observations are available for the target star, but for the purpose of microlensing modeling it is sufficient to consider data within the 2006 season. In fact, the peak was reached on 2006 July 15 (HJD' $\sim$ 3931.64) and the full duration of the microlensing magnification is about 50 days, fully confined well within the season. Therefore, we are left with 738 data points in the MOA-R broad band, which spans across Johnson-Cousins-Bessell R and I bands. Unfortunately, no MOA-V data are available for this event, a fact that makes a color analysis of the source particularly difficult. In addition, there is no data from the Optical Gravitational Lensing Experiment (OGLE; \citep{OGLE2003}), the other main survey active at the time of the event, because the event fell at the edge of the field.

As customary in microlensing field \citep{Miyake2012}, error bars are adjusted as $\sigma_i^2=k\sqrt{\sigma_{i,orig}^2+e_{min}^2}$, where $e_{min}$ is fixed by requiring that the residuals are similar for high-magnification and low-magnification sections of the light curve, while $k$ ensures that $\chi^2/d.o.f.=1$ for the best model. 

\section{Modeling} \label{sec:models}

The microlensing effect depends on a special angular scale called Einstein angle or angular Einstein radius (cfr. \citet{Gaudi2012} for a review):
\begin{equation}
\theta_E=\sqrt{\frac{4GM}{c^2}\frac{D_{LS}}{D_{OL}D_{OS}}},
\end{equation}
where $G$ is the Newton constant, $c$ the speed of light, $M$ is the total lens mass, $D_{OL}$ is the lens-observer distance, $D_{LS}$ is the lens-source distance, $D_{OS}$ is the observer-source distance. The magnification of the source flux reaches a maximum when the closest approach between the lens and the line of sight to the source is reached and then declines afterwards. The light curve of a basic model with one lens and one source (model 1L in the following) just depends on four parameters: the time of lens-source closest approach $t_0$, the impact parameter in units of the Einstein radius, $u_0$, the Einstein time $t_E=\theta_E/\mu$, where $\mu$ is the relative lens-source proper motion, and the ratio of the source angular radius to the Einstein radius $\rho_*=\theta_*/\theta_E$. The latter is only measured if $u_0$ is comparable to $\rho_*$, otherwise only an upper limit is obtained. For the brightness profile of the source we adopt linear limb darkening with coefficient $a_{MOA}=0.508$, as explained in Section \ref{sec:source}. In addition to these parameters, we also have the source flux $F_*$ and a possible blending flux $F_B$, coming from stars that are undistinguished from the source at the resolution of the observations. 

\begin{figure}[ht!]
\epsscale{0.75}
\plotone {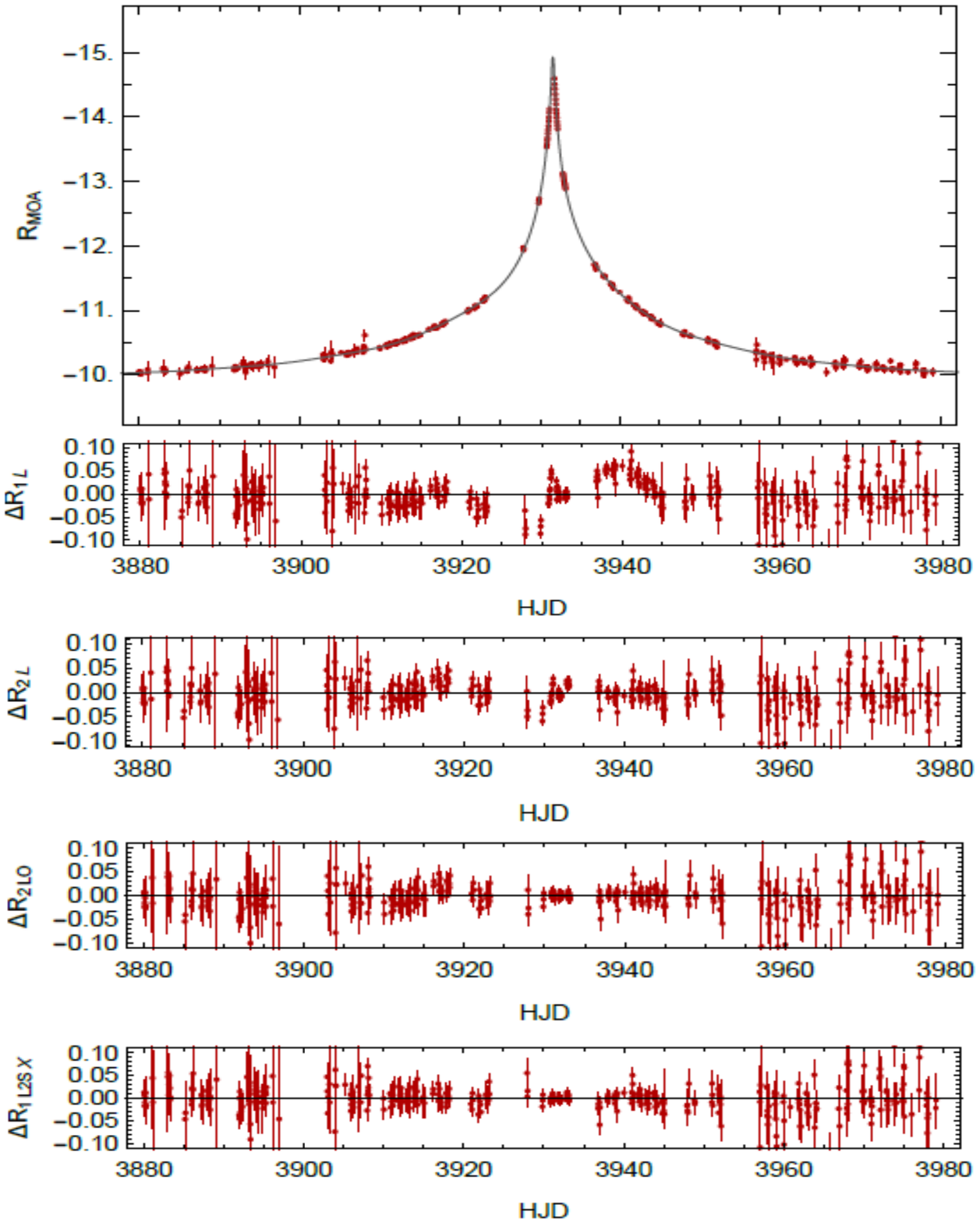}
\caption{Light curve of MOA-2006-BLG-074 with the best model 1L2SX. We also show the residuals for the single-lens model 1L, the static binary lens model 2L, the binary lens model including orbital motion 2LO, the single lens binary source model 1L2SX. }\label{fig:model}
\end{figure}

\begin{figure}[ht!]
\epsscale{0.75}
\plotone {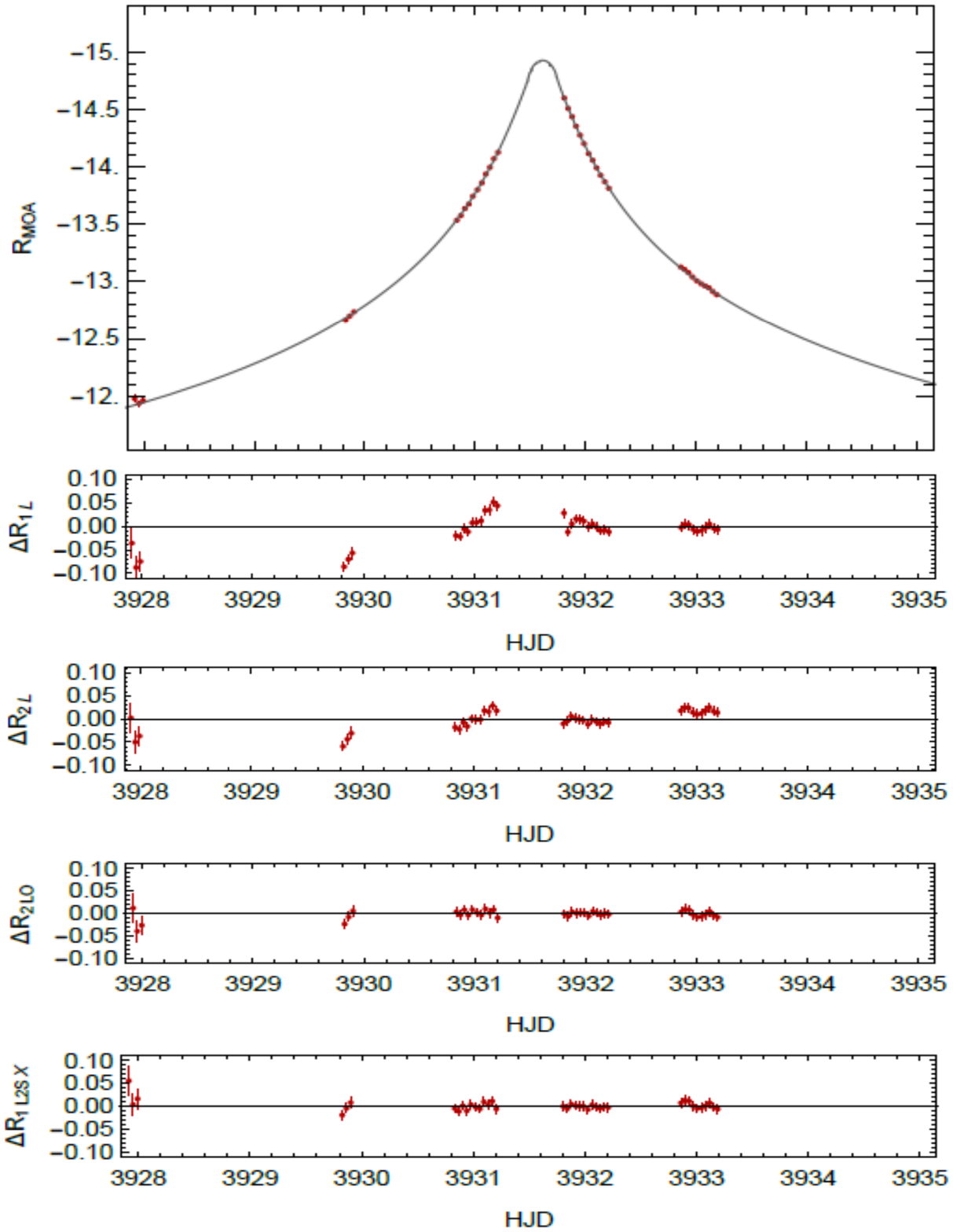}
\caption{Zoom of the light curve of MOA-2006-BLG-074 with the best model. We also show the residuals for the single-lens model 1L, the static binary lens model 2L, the binary lens model including orbital motion 2LO, the single lens binary source model 1L2SX.
}\label{fig:zoom}
\end{figure}

In Fig. \ref{fig:model} we see that the residuals of the observed lightcurve from such model are evident in the peak and after the peak around HJD$\sim$3940. The anomalies at the peak suggest the possibility that the central caustic is not point-like due to the distortion of a secondary lens. 

\subsection{Binary lens models}

In the retrospective analysis carried by MOA collaboration, MOA-2006-BLG-074 was independently identified as a possible planetary event using three different modeling codes \citep{Sumi2010,Bozza2010,Bennett2010}. All of them converged to the same close binary model with a mass ratio of $q \sim 10^{-4}$, which clearly identified this as a promising planetary candidate deserving a deeper investigation.

A binary lens model is minimally described by three additional parameters: the planet to host mass ratio $q$, the planet to host angular separation in units of the Einstein radius $s$, and the angle between the planet-host axis and the source trajectory $\alpha$. In order to explore the parameter space, we use the RTModel platform\footnote{http://www.fisica.unisa.it/GravitationAstrophysics/RTModel.htm}, which is based on a template library approach \citep{MaoDistefano1995}. The basic magnification calculation is performed by the {\textit VBBinaryLensing} code \citep{Bozza2010,Bozza2018,Bozza2020}. With this approach, we find two planetary solutions with $s<1$, and their corresponding wide duals with $s\rightarrow 1/s$ \citep{Dominik1999,GriestSafizadeh1998}. Among these solutions, the best one plotted in Figs. \ref{fig:model} and \ref{fig:zoom} as model $2L$ is the wide one with the smaller separation. Its parameters, listed in Table \ref{table:tab1}, indicate a planet with $q\sim 10^{-4}$ in a nearly resonant configuration. As anticipated, the $\chi^2$ of the binary lens model is much lower than that of the single-lens model. This is confirmed by the fact that the residuals around $HJD=3940$ in Fig. \ref{fig:model} are flattened. However, some residuals remain in the peak (Fig. \ref{fig:zoom}). We also note that the alternative model with slightly wider separation flattens the residuals at the peak while leaving the deviation at $HJD=3940$ untouched. The close duals of these two models have higher $\chi^2$ and have a poorer performance.

\begin{deluxetable*}{cccccc}
\tablecaption{Parameters for all models considered.}\label{table:tab1}
\tablewidth{0pt}
\tablehead{
 & (Unit) & 1L & 2L & 2LO & 1L2SX 
}
\startdata
$t_{E}$ &  days & $40.12 ^{+0.55}_{-0.47}$ & $39.84^{+0.46}_{-0.54}$ & $40.86^{+1.05}_{-0.91}$ & $33.29^{+0.31}_{-0.31}$  \\
$t_0$ & $HJD$ & $3931.627^{+0.001}_{-0.002}$ & $3931.627^{+0.002}_{-0.001}$ & $3931.637^{+0.001}_{-0.006}$ & $3931.610^{0.001}_{-0.002}$  \\
$u_0$ & &$0.0102^{+0.0002}_{-0.0003}$ & $0.0084^{+0.0003}_{-0.0002}$ & $-0.0067^{+0.0004}_{-0.0005}$ & $0.0091^{+0.0001}_{-0.0002}$ \\
$\rho_{*}$ & & $0.01192^{+0.00026}_{-0.00043}$ & $0.00942^{+0.00037}_{-0.00024}$ & $0.00633^{+0.00110}_{-0.00041}$  & $0.00985^{+0.00064}_{-0.00022}$\\
$\alpha$ & & - & $3.133^{+0.001}_{-0.001}$ & $-3.130^{+0.032}_{-0.005}$ & - \\
$s$ & & - & $1.088^{+0.002}_{-0.001}$ & $0.755^{+0.015}_{-0.011}$  & -   \\
$q$ & &- & $0.00014^{+0.00001}_{-0.00001}$ & $0.00063^{+0.00005}_{-0.00013}$ & -   \\
$\pi_{E,N}$ & &- & - & $-0.473^{+1.053}_{-0.177}$ & -   \\
$\pi_{E,E}$ & &- & - & $0.323^{+0.069}_{-0.352}$ & -  \\
$\gamma_{\parallel}$ & days$^{-1}$ &- & - & $0.0483^{+0.0030}_{-0.0020}$ & -  \\
$\gamma{\perp}$ & days$^{-1}$& - & - & $-0.0060^{+0.0030}_{-0.0030}$ & -  \\
$\gamma_{z}$ & days$^{-1}$& - & - & $0.0389^{+0.0031}_{-0.0004}$ & - \\
$\xi_{\perp}$ & & - & - & - & $0.0198^{+0.0018}_{-0.0017}$  \\
$\xi_{\parallel}$ & & - & - & - & $0.0088^{+0.0007}_{-0.0001}$  \\
$\omega$ & days$^{-1}$ & - & - & - & $0.442^{+0.001}_{-0.008}$  \\
$i$ & &- & - & - & $-0.054^{+0.017}_{-0.049}$ \\
$\phi$ & & - & - & - & $4.493^{+0.630}_{-0.049}$ \\
$q_{s}$  & & - & - & - & $<0.4498$   \\
\hline
$R_{base}$ & mag & $-9.913^{+0.002}_{-0.003}$ & $-9.921^{+0.003}_{-0.003}$ & $-9.924^{+0.008}_{-0.003}$ & $-9.937^{+0.009}_{-0.003}$ \\
$g$ & & $0.647^{+0.022}_{-0.017}$ & $0.710^{+0.020}_{-0.023}$ & $0.782^{+0.051}_{-0.032}$  & $0.422^{+0.026}_{-0.004}$  \\
$\chi^2$ & &  1594.4 & 849.5 & 713.1 & 699.1 \\
\enddata
\end{deluxetable*}

The basic binary model just described assumes a static lens and approximates the relative source-lens motion as rectilinear. Such static models only work if the timescale of the microlensing event is much smaller than the orbital period of the binary lens and the annual motion of the Earth around the Sun. It is natural to conjecture that the remaining discrepancies between the static model and the data is due to these two effects that have been neglected in model 2L.

The annual parallax caused by the Earth revolution around the Sun is parameterized by the two components $\pi_{E,N}$ and $\pi_{E,E}$ of the parallax vector along the North and East directions in the sky. The module of this vector is
\begin{equation}
\pi_E=\frac{\pi_{rel}}{\theta_E}, \pi_{rel}=\frac{au}{D_{OL}}-\frac{au}{D_{OS}},
\end{equation}
and its direction is given by the proper motion of the lens with respect to the source \citep{Gould2000}. 

The second effect is due to the orbital motion of the planet with respect to the host star. A complete description of the orbital motion would require the three components of the companion's velocity relative to the host star, the projection of the separation of the binary lens along the line of sight in units of $\theta_E$ and the orbital semimajor axis or another related quantity. In general, however, the relatively short window opened by microlensing only allows the measurement of two components of the projected angular velocity of the planet, $\gamma_{\parallel}= (ds/dt)/s$ and $\gamma_{\perp}=-(d\alpha/dt)$ at the reference time $t_0$. In many cases, it is sufficient to approximate the orbital motion by using constant values of these two quantities  \citep{Hwang2010b}. However, since such approximation does not correspond to a physical orbital motion, we prefer to study models including the component of the angular velocity along the line of sight $\gamma_{z}= (ds_z/dt)/s$ \citep{Skowron2011,Bozza2020}. With the assumption of circular orbital motion, these three parameters are sufficient to completely characterize the orbit and thus allow us to explore a sub-space of possible physical solutions, as opposed to the two-parameter linear orbital motion. As stated before, the additional velocity parameter is very poorly constrained by typical microlensing events. Only in a few exceptional cases a full Keplerian motion had to be explored \citep{Wyrzykowski2020}. We refer to the binary model including the parallax effect and the circular orbital motion as model 2LO. 

In all the cases, the parameter space is significantly enlarged. In order to lead the broadest exploration possible, we start from the static solution and minimize the $\chi^2$ by a Levenberg-Marquardt run. After that, we start Markov chains at large temperatures and record all separate minima we find. For each of this minima we run chains at lower temperature until we single out the best solution. In the case of annual parallax, the symmetry of the model for reflection around the star-planet axis is broken and we have to consider possible reflections separately. 

As results of our search, we first note that all models with parallax and no orbital motion for the lens converge to unphysically large values of the parallax, with $\pi_E>2$, which would imply a very close-by or very small lens. For this reason we discard such models and take this outcome as a suggestion that some motion with shorter timescale is at work in this event.

\begin{figure}[ht!]
\plotone{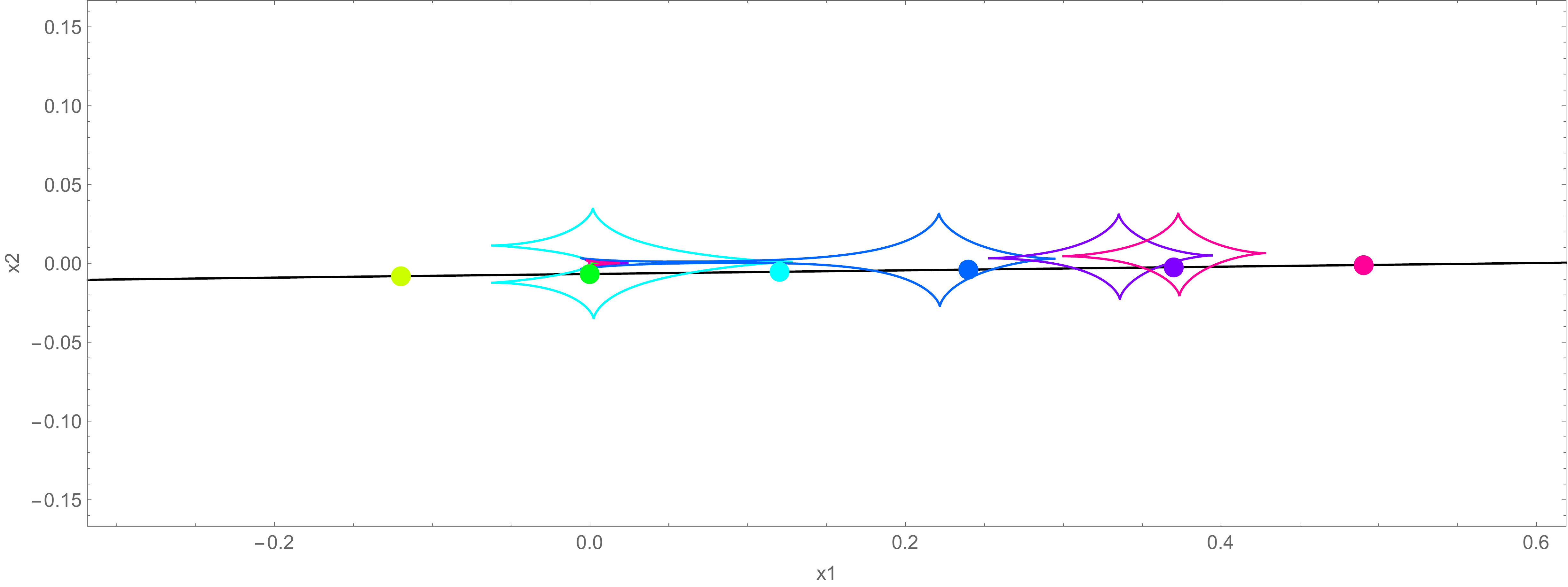}
\caption{Caustic configuration for the binary orbital model $2LO$ presented in Table \ref{table:tab1}, with the source moving along the straight line from left to right. The green color refers to time $t_0$, while other colors represent the source and caustics in time-steps of 5 days.
\label{fig:caustic}}
\end{figure}

A more satisfactory model (named $2LO$ in Figs. \ref{fig:model},\ref{fig:zoom} and Table \ref{table:tab1}) is obtained when the lens orbital motion is included. The $\Delta \chi^2=136$ with respect to the static model $2L$ can be appreciated especially in the peak region visible in Fig. \ref{fig:zoom}. Some deviation is left on the left side at $HJD\sim 3917$. The parallax components of this model are poorly constrained and still compatible with zero at $1\sigma$, which confirms that parallax is not the main motion to be considered here. The orbital motion has a zero $\gamma_\perp$ component, which means that the planet orbit is seen edge-on. Note also that the value of $s<1$ indicates that this model comes as the evolution of one of the static close models. However, the high value for $\gamma_\parallel$ warns that the separation rapidly evolves from the close to the wide regime. This evolution is apparent in Fig. \ref{fig:caustic}, where we see that the caustic is still in the close topology when the source passes close to the central caustic. Soon after, the caustic becomes resonant and finally the planetary caustic detaches to the right. The impression is that in order to explain the peak anomaly together with the wing anomalies the caustic ``follows'' the source along its motion, which looks quite suspicious.

The  orbital period for this solution is about 161 days, which could seem reasonable at a first glance, but we will delay the discussion of its implication on the masses of the system after the analysis of the source in Section \ref{sec:bayes}.

\subsection{Binary source models}

In parallel with the binary lens model, we test an alternative direction to explain the anomalies seen in Fig. \ref{fig:model}. Instead of adding a second lens, we add a second source \citep{GriestHu1992}. From the very beginning we include the orbital motion of the two sources around the common center of mass under the simplifying hypothesis of a circular trajectory. Therefore, in addition to the 4 parameters of the 1L model ($t_E$,$t_0$,$u_0$, $\rho_*$) we introduce 5 more parameters \citep{RahvarDominik2009}: the inclination of the orbital plane $i$, the phase from the ascending node $\phi$, the angular velocity $\omega$ and the two projections of the node line parallel and perpendicular to the source velocity at time $t_0$: $\xi_\parallel$ and $\xi_\perp$, with the angular orbital radius being $\xi=\sqrt{\xi_\parallel^2+\xi_\perp^2}$ in units of $\theta_E$. Finally, we consider the possibility that the secondary source contributes to the observed flux. Therefore, we introduce the mass ratio $q_s$ of the secondary to the primary. For simplicity, we assume a power-law mass-luminosity relation $F_2=F_* q_s^4$ and a mass-radius relation $\rho_{2}=\rho_* q_s^{0.8}$ \citep{CarrollOstlie2006}. We will see that the results do not depend on the particular choice of these relations.

We have started our search in the parameter space from the best single-lens-single-source model. The 5 orbital parameters have been set to zero in the initial condition, while $q_s$ has been set to $0.1$. Similarly to model $2LO$, we have run Markov chains with decreasing temperature, and branched different chains for independent provisional minima. At the end of our search, our best solution (labeled as $1L2SX$) improves the fit with respect to model $2LO$ by $\Delta \chi^2=14$. This is the light curve plotted in Figs. \ref{fig:model},\ref{fig:zoom}, which fits the peak region well, while performing much better on the wing anomalies.

\begin{figure}[ht!]
\plotone{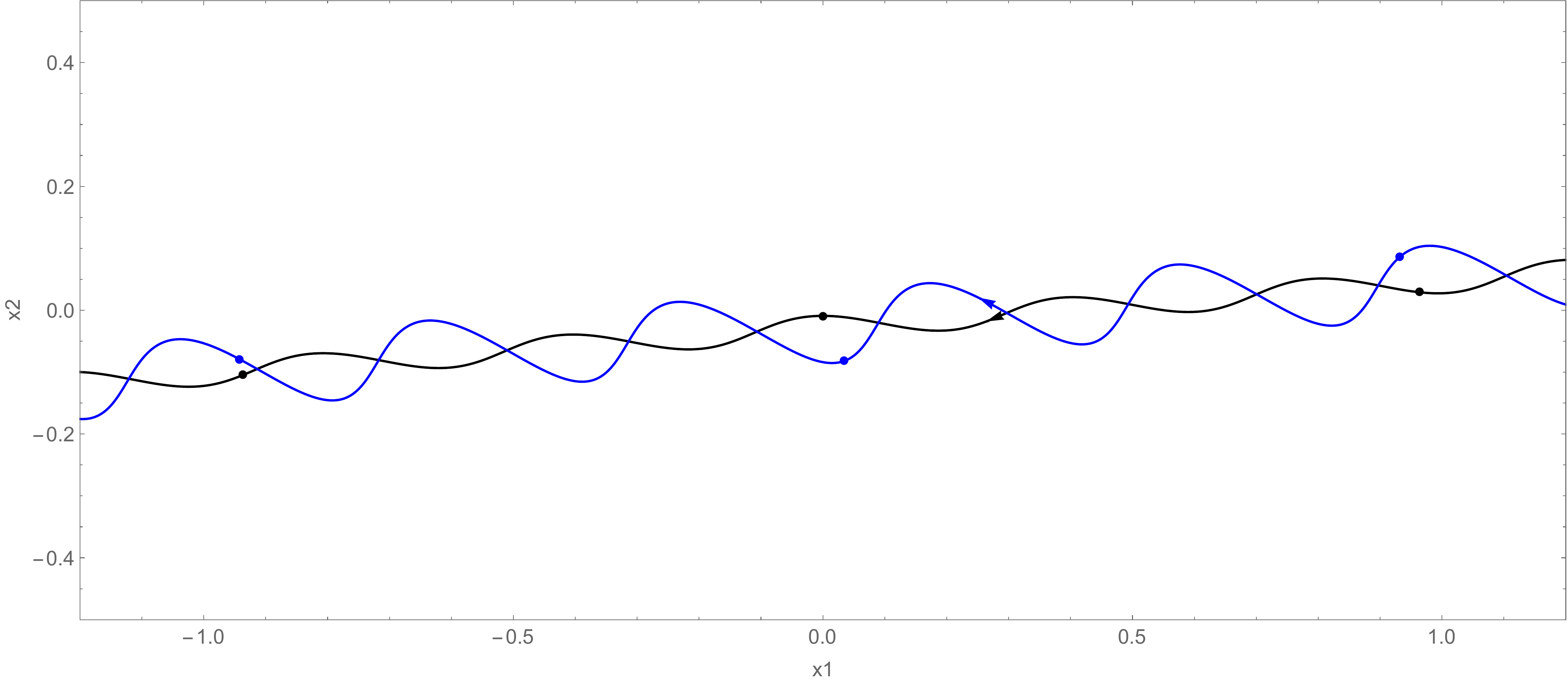}
\caption{Trajectory of the primary source (black) and secondary source(blue) for the best model with one lens and two sources in Table \ref{table:tab1} with a choice of the mass ratio $q_s=0.33$. We have marked the positions of the sources at $t_0-t_E,t_0,t_0+t_E$.
\label{fig:binarysource}}
\end{figure}

Before acclaiming this binary source as the correct model, we must go through all possible checks that this solution is indeed physically acceptable. The orbit we find is nearly face-on with a period of $P=14.2d$. The orbital radius for the primary is $\xi=0.022$ in units of the Einstein angle, which suggest that we are dealing with either a close binary system or a system in which the secondary is much lighter than the primary. Indeed, for the mass ratio we only find an upper limit $q_s<0.422$, which indicates that the secondary basically intervenes through the reflex motion of the primary, while its contribution to the total flux is not essential. Also for this model we delay a full discussion of the physical constraints after the source analysis. Fig. \ref{fig:binarysource} shows the trajectories of the two sources after choosing a mass ratio $q_s=0.33$ (see Section \ref{sec:bayes}).

\subsection{Other models} 

Both model $2LO$ and especially $1L2SX$ provide a good fit to the observations. Nevertheless, in this subsection we mention the performance of other more complicated models that are worth checking.

First we consider a static triple-lens model $3L$. We have already mentioned that for a static binary model we have found two solutions with a close-in planet and two solutions with a planet in a wide configuration. Therefore, we have considered having two planets in the system placed in two of the positions of these four different binary solutions, taking all six possible combinations into account. We have found that the best model arises with two planets in the wide configuration. However, the corresponding $\chi^2= 729.4$ is not better than that for the model $2LO$. Since the models $2LO$ and $1L2SX$ provide a better explanation to the data with a lower number of bodies, there is no reason to consider a triple lens model further.

Another possibility is that we have two sources and two lenses at the same time. We may combine the parameters of the binary lens model $2L$ and the binary source model $1L2SX$ to improve the $\chi^2$ further. Indeed we find a model with $\chi^2=674.9$ sharing the characteristics of the two parent models. However, the $\Delta \chi^2 =24.2$ with respect to the $1L2SX$ is far below any conventional threshold in microlensing observations to claim evidence of an additional body and can be explained by overfitting of systematics. 

Finally, we have also checked that the inclusion of annual parallax in the $1L2SX$ only leads to a very modest improvement of the $\chi^2$.

\section{Source analysis} \label{sec:source}

At this stage we are left with two competing models: model $1L2SX$ with one lens and two sources and model $2LO$ with two lenses and one source. We must discriminate between them on the basis of physical constraints. In order to do that, we can use the information on the source size, which is well constrained in both cases, to derive the size of the angular Einstein radius $\theta_E$. In this section we will give all the values for model $1L2SX$ and quote the corresponding values for model $2LO$ in round brackets. 

The source size parameter is $\rho_*= 9.85 \times 10^{-3}$ ($6.33\times 10^{-3}$ for model $2LO$).\footnote{Note that the values of $\rho_*$ for the two models are quite different. However, the two Einstein times are in a reverse relation, so that the source crossing time $t_*=\rho_* t_E$ is in marginal agreement at $2\sigma$. The slightly shorter $t_*$ for model $2LO$ can be understood as a consequence of the finite extension of the central caustic perturbed by the secondary body.} If we are able to fix the angular size of the source, from this parameter we can derive the Einstein angle, which provides the lensing scale, constraining the mass and distance to the lens. 

\begin{figure}[ht!]
\epsscale{0.65}
\plotone{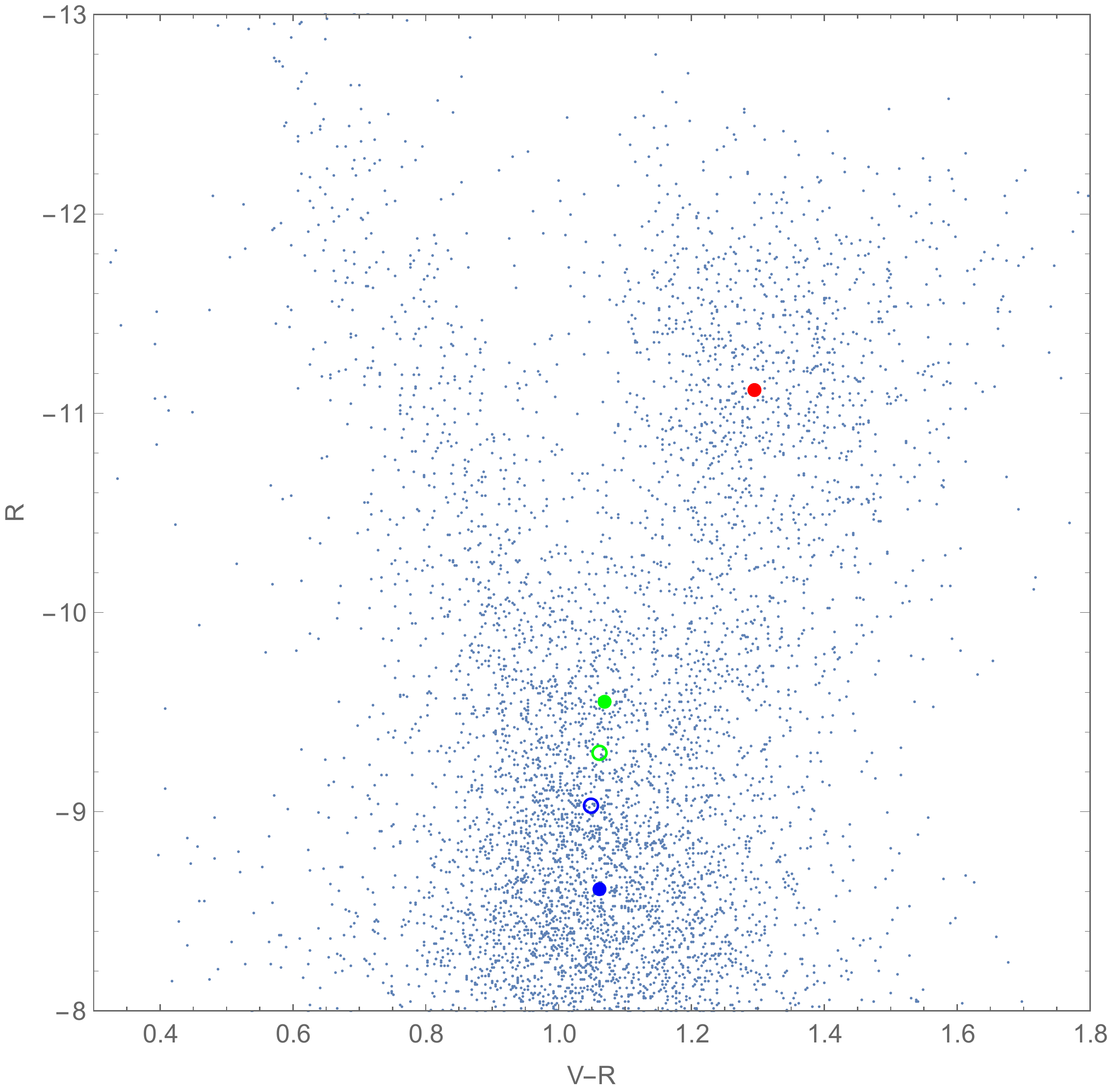}
\caption{Color magnitude diagram (CMD) of the stars in the field of MOA-2006-BLG-074. The red dot shows the position of the red clump. The green dot represents the source star and the blue dot shows the blend position for model $1L2SX$. The empty circles represent the same objects for the $2LO$ case. \label{fig:cmd}}
\end{figure}

Unfortunately, MOA observations are only available in MOA-R band for this event, and no additional observations have been taken by other telescopes during the microlensing event. In Fig. \ref{fig:cmd} we show a color-magnitude diagram for the MOA field including our microlensing event. We can easily find the centroid of the red clump at $R_{MOA,clump}=-11.120\pm0.009$ and $(V-R)_{MOA,clump}=1.295\pm0.006$. In order to convert MOA magnitudes to standard Johnson-Cousins bands, we cross-calibrate with OGLE-III photometry \citep{Szymanski2013} and derive the following relations:

\begin{eqnarray}
I_{OGLE}=R_{MOA}+28.206-0.217(V_{MOA}-R_{MOA}) \pm 0.002&&\\
V_{OGLE}=V_{MOA}+28.510-0.146(V_{MOA}-R_{MOA}) \pm 0.002 && 
\end{eqnarray}

We then obtain $I_{OGLE,clump}=16.805\pm0.009$ and $V_{OGLE,clump}=18.496\pm0.011$. Comparing with the intrinsic red clump color of $(V-I)_{clump,0} = 1.06\pm0.07$  \citep{Bensby2011} and intrinsic magnitude $I_{clump,0} = 14.44\pm0.04$ \citep{Nataf2013}, we derive an extinction $A_I = 2.36 \pm 0.04$ and a reddening $E(V - I) = 0.63 \pm 0.07$.

Our models are characterized by blending parameters $g=0.421 (0.782)$ on a baseline of $R_{base}=-9.931 (-9.924)$. From this we derive a source magnitude $R_{MOA,*} = -9.549 (-9.297)\pm 0.010$. Without any observations in the MOA-V band, we can just make a minimal hypothesis for the source color. In fact, the levels $R_{MOA,*}$ for the two models in Fig. \ref{fig:cmd} correspond to the region of the turn-off point of the main sequence. The range of colors of the stars in the field is relatively narrow. So, we assume that the source color can be represented by the average color of the stars in the field at $R_{MOA,*}$ with an uncertainty given by the root-mean-square. Of course, this uncertainty will propagate to the final measure of the source radius. We obtain $V_{MOA,*} = -8.480 (-8.235)\pm 0.164 (0.155)$. These values then convert to $I_{OGLE,*}=18.424 (18.678)\pm0.036 (0.034)$ and $V_{OGLE,*}=19.874 (20.120)\pm0.139 (0.133)$, which correspond to de-reddened values for the source of 
$I_{OGLE,*,0}=16.064 (16.318)\pm 0.054 (0.052) $ and $V_{OGLE,*,0}=16.884 (17.130)\pm0.160 (0.155)$.

At this point, we use \citet{BessellBrett1988} to transform V-I to V-K color and then use the empirical formula from \citet{Kervella2004}, to find the source angular radius $\theta_*=2.15 (1.89) \pm 0.56 (0.47) \mu$as. Note that the higher blending for model $2LO$ leads to a slightly smaller source. However, the $\rho_*$ is also smaller for this model. We can thus estimate the angular Einstein radius:

\begin{equation}
\theta_E= \frac{\theta_{*}}{\rho_{*}}= 0.22 (0.30)\pm 0.06 (0.10) \, mas
\end{equation} 
 and the lens-source relative proper motion $\mu_{rel}=\theta_E/t_E= 2.40 (2.67) \pm 0.26 (0.28) $ mas $yr^{-1}$. 
 
Since the finite source effects are relevant in both models, we have included the limb-darkening of the source brightness profile in our modeling. In order to estimate it correctly we proceed in the following way: first we take the value of$(V-I)_{OGLE}$=1.450 and the magnitude $M_I=3.96$ of the source, then we simulate a stellar population with solar metallicity using IAC-STAR \citep{Aparicio2004} with the \citet{Bertelli1994} stellar evolution library and \citet{CastelliKurucz2003} bolometric correction library to obtain $\log g = 4.31$ and $T_{eff}=5625 K$. From these values we get the linear limb darkening coefficients in I and R band $a_I=0.462$ and $a_R=0.554$. Since the MOA-R band extends on both R and I band almost equally, we use the mean value $a_{MOA}=0.508$ \citep{Kondo2019}. All models presented in the previous session have been calculated with this limb darkening coefficient.

\section{Physical constraints} \label{sec:bayes}

In this section we will implement all remaining physical constraints once the angular Einstein radius has been fixed according to the source analysis in Section \ref{sec:source}. Since we do not have a parallax measure in either models, we resort to a Bayesian analysis for the estimate of the mass and distance of the lens. We assume the galactic model of \citet{Dominik2006} to fix our prior probabilities and use the measured $t_E$ and $\rho_*$ to constrain the lens physical parameters. We also include the constraint from the lens flux, for which we use the mass-luminosity model of \citet{CastelliKurucz2003}, to estimate the magnitude $R_{MOA}$ of the lens and impose that it does not exceed the blend flux. For the source distance, the Galactic model predicts a modal value of $D_{OS}=9$ kpc for the line of sight that we will adopt in the following.

Since all values for the parameters are slightly different for model $2LO$ and $1L2SX$, we present the results of this analysis separately. 

\subsection{Constraints on the binary lens model}

\begin{figure}[ht!]
\epsscale{0.55}
\plotone{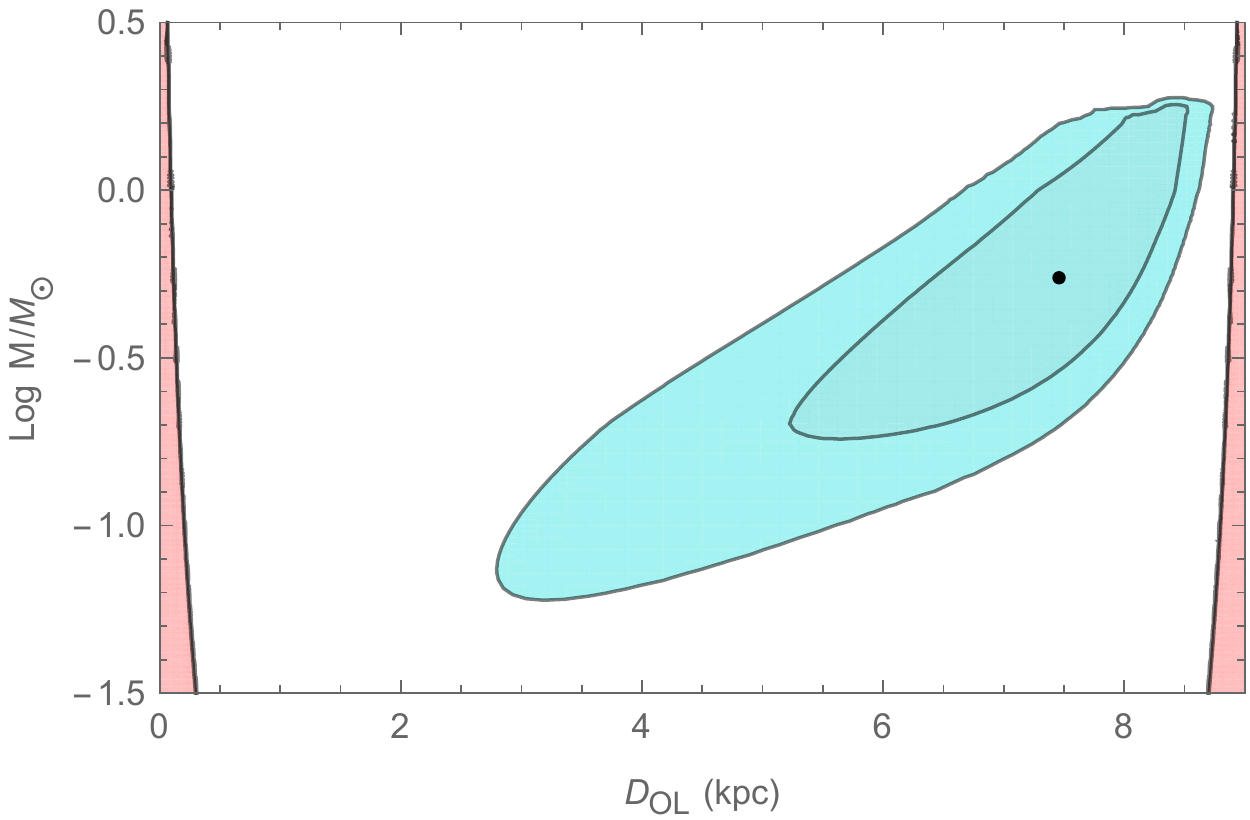}
\caption{Allowed regions (in blue) in the plane $D_{OL}$(kpc) - Log M/$M_{\sun}$ at $68\%$ and $95\%$ CL for the $2LO$ model. In red, the regions allowed by the Keplerian constraint.  }\label{fig:BayesL}
\end{figure}

Fig. \ref{fig:BayesL} shows the posterior probability for model $2LO$. A slight preference for a $0.5 M_\sun$ lens in the bulge arises, which would be consistent with the absence of annual parallax.

However, the mass and the distance to  the lens must be compatible with the orbital period derived in our model. Following \citet{Skowron2011}, in order to have a bound system, the projected kinetic energy must be less than the projected potential energy. This bound translates to
\begin{equation}
\frac{v^2_{\perp}r_{\perp}}{2GM}=\frac{(\gamma^2_{\perp}+\gamma^2_{\parallel})s^3\theta^3_ED^3_{OL}}{2GM}<1.
\end{equation}

Unfortunately, in the plane $(D_{OL},M)$ this constraint selects two extremely thin slices very close to the observer or very close to the source, as we can see in Fig. \ref{fig:BayesL}. There is no overlap between the blue and red regions. The very fast orbital motion needed to justify all anomalies is possible only in the very unlikely configurations of a lens at a few pc away from the source or from the observer.  

\subsection{Constraints on the binary source model}

\begin{figure}[ht!]
\epsscale{0.55}
\plotone{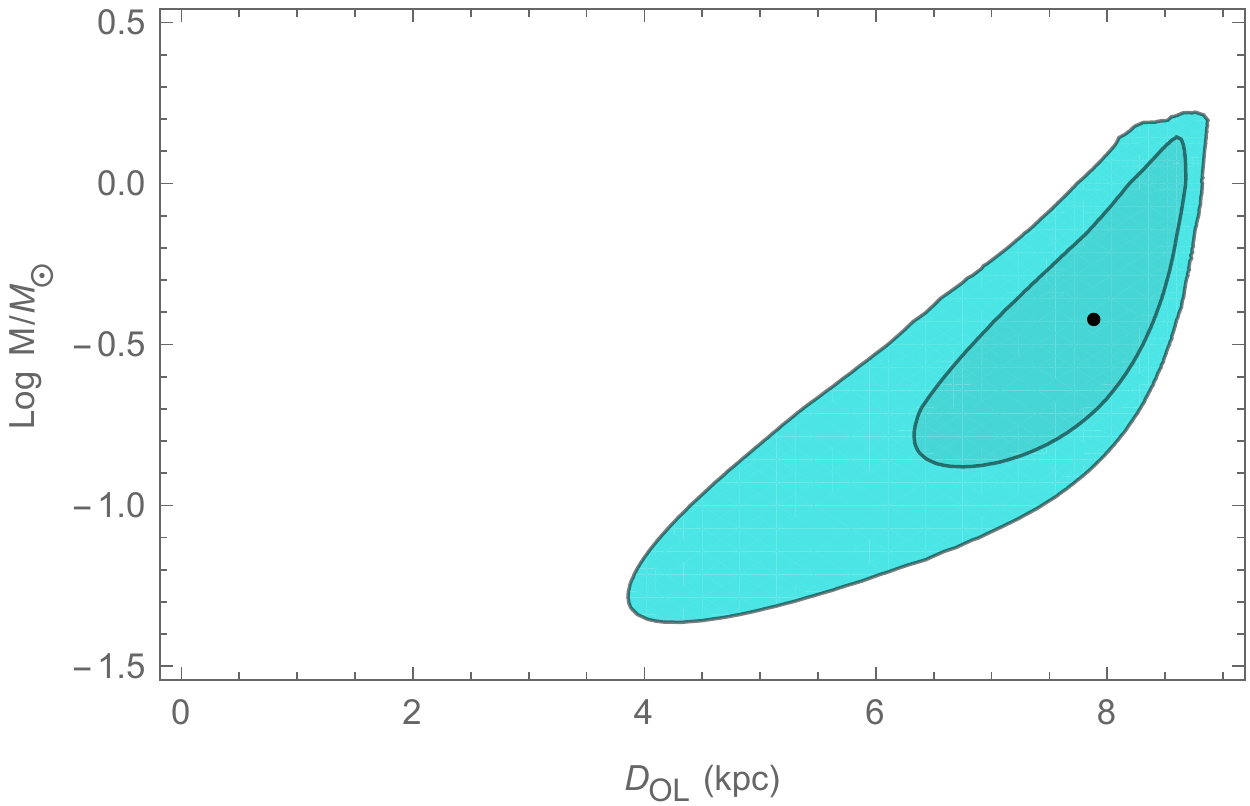}
\caption{Allowed region in the plane $D_{OL}$(kpc) - Log M/$M_{\sun}$ at $68\%$ and $95\%$ CL for model $1L2SX$. } \label{fig:BayesS}
\end{figure}

For model $1L2SX$ we obtain the posterior probability in Fig. \ref{fig:BayesS}, in which the lens is lighter and even closer to the source with respect to model $2LO$. We get $M_L = 0.38\pm0.23 M_\sun$, $D_{OL}= 7.9^{+0.6}_{-1.0}$ kpc. The lens has a $70\%$ probability to be a bulge star rather than a disk star.

From the analysis of the source flux in Section \ref{sec:source}, besides the radius of the source and its limb darkening coefficient, we can also estimate its mass by comparing to the stellar libraries. Taking into account the uncertainties discussed therein, we get $M_S=1.32 \pm 0.36 M_\sun$. With this information, we can finally test the consistency of the xallarap solution with the third Kepler's law \citep{HanGould1997,Miyake2012}. In logarithmic form this reads
\begin{equation}
    f_K(q_s,\xi,\omega) = K, \label{Keplersource}
\end{equation}
where
\begin{equation}
    f_K(q_s,\xi,\omega)=\log\left[\frac{\omega^2 \xi^3 (1+q_s)}{q^{3}_{s}}\right],
\end{equation}
\begin{equation}
    K=\log\left[\frac{M_1 G}{(D_{OS} \theta_E)^3} \right]=-14.16\pm0.90.
\end{equation}

All quantities in $f_K$ are fitting parameters of model $1L2SX$, while the quantities in $K$ have been estimated by use of the Galactic model and stellar libraries. Only the models whose fitting parameters satisfy this constraint within the uncertainties can be considered as physically allowed.

\begin{figure}[ht!]
\epsscale{0.55}
\plotone{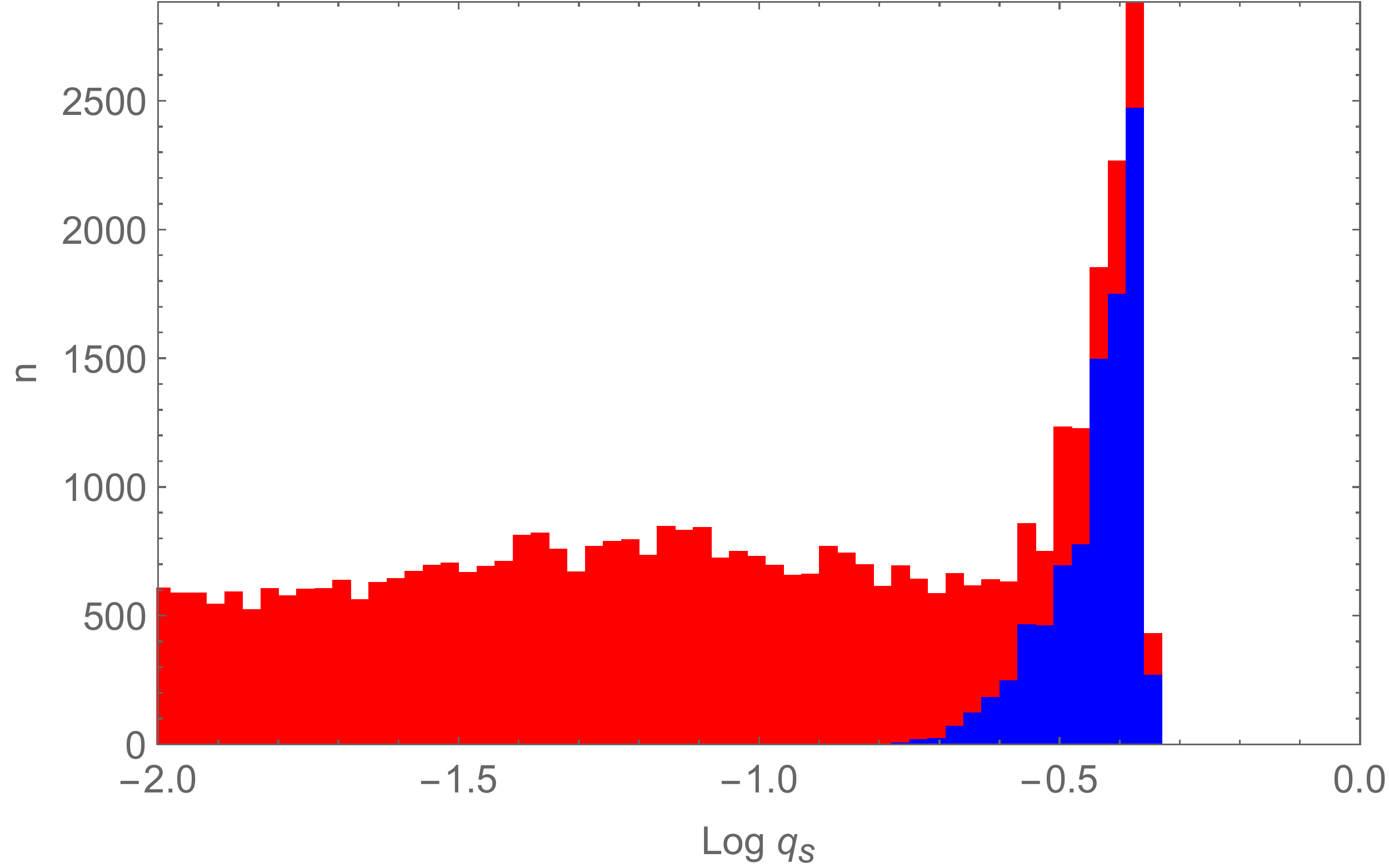}
\caption{Distributions of the parameter $\log q_S$. The red histogram is obtained with a flat prior, while the blue one is the same distribution where each point in the Markov chain is weighted by the prior $P(q_s,\xi,\omega)$ in Eq. (\ref{prior}).  }\label{fig:qs}
\end{figure}

Fig. \ref{fig:qs} shows the distribution in red for $\log q_s$ generated by a Markov chain with a flat prior in this parameter. We already see that the distribution is generally flat but then has a sharp peak on the right end for higher mass ratios before dropping to zero when the source becomes too luminous to be compatible with the observed light curve. The blue histogram is obtained after weighting each point in the Markov chain by the Gaussian prior
\begin{equation}
    P(q_s,\xi,\omega)=\exp\left[-\frac{(f_K(q_s,\xi,\omega)- K)^2}{2\sigma_K^2} \right], \label{prior}
\end{equation}
where  $\sigma_K$ is the uncertainty in the combination $K$. Notably, the red peak and the blue peak coincide, which means that the likelihood itself already favors the same mass ratio that is selected by the Keplerian constraint. Since the mass ratio only intervenes when the light of the secondary source becomes relevant, this can be interpreted as a marginal detection of the companion to the main source. Such detection is perfectly compatible with the Keplerian constraint derived from the properties of the primary source. In Table \ref{table:tab2} we summarize the parameters of the binary system acting as the source.

\begin{table*}
\centering
\begin{tabular}{lc}
\hline \hline
$D_{OL} (kpc)$     &  $7.9^{+0.6}_{-1.0}$ \\
$M_L (M_\sun)$     &  $0.38\pm0.23$ \\
$M_{S1} (M_\sun)$     &  $1.32 \pm 0.36$ \\
$M_{S2} (M_\sun)$     &  $0.44 \pm 0.14$\\
$a (au)$     &  $0.043 \pm 0.012$\\
$ P (d) $ & $14.2 \pm 0.2$ \\
$i (^\circ)$ & $-3.0 \pm 1.9$ \\
 \hline
\end{tabular}
\caption{Physical parameters for model $1L2SX$.}\label{table:tab2}

\end{table*}

\section{Discussion and conclusions} \label{sec:disc}

In the retrospective analysis of MOA microlensing events a sizeable number of candidate planets has been found. For many of these events there are no additional observations that may complement the MOA dataset. For the event analyzed in this paper, MOA-2006-BLG-074, we only have observations in a single band, without any possibility to obtain color information. Nevertheless, modeling of the available photometry is possible and leads to interesting results that can be useful to drive this and future analysis in similar situations.

MOA-2006-BLG-074 is a high-magnification event ($A_{max}\sim 110$) with a distinct anomaly on the peak. Such anomaly can be naturally interpreted as the result of a central caustic perturbed by a small planet. However, the presence of modulations on the wings indicates that higher order effects must be taken into account. Unfortunately, the required orbital motion of the hypothetical planet is so fast that the Keplerian constraint for a bound system can only be fulfilled by a lens in the immediate neighborhood of the source.

The alternative possibility of a binary source and a single lens is able to explain the peak anomaly and the wing modulations with an orbital motion of 14 days. Not only this model is favored by a lower $\chi^2$, but it also leads to a physically viable solution in which the primary is a star at the turn-off point of the main sequence and the secondary is a K-M dwarf. The light of the secondary may have been marginally detected. Observations in a second band would have been very useful to check for a color difference of the two sources. The lens is probably an M-dwarf in the bulge. The blending light could be explained by the lens if this is at the brighter end of the allowed region, but it is probably another star unrelated with the microlensing event. Note that 15 years has passed since the detection of this event: the lens and the source should be separated by 36 mas, which is still too low to be resolved by current facilities \citep{Bennett2006}, also considering the faintness of the source.

The microlensing event analyzed in this paper has not led to the discovery of a new planet in spite of being selected by three different modeling platforms as a promising candidate. Although binary sources have long been studied as possible contaminants in planetary microlensing searches, they are often overlooked in real-time modeling or in the selection process from large datasets. In view of the incoming Roman Galactic Exoplanet Survey \citep{Penny2019}, the contamination by binary sources has never been seriously estimated. Part of the reason is that it is difficult to find realistic estimates of the fraction of binary sources in the bulge, their separations and their orbital periods. Yet, it would be interesting to check how often the expected planetary signal can be mimicked by a competing binary source model. In general, distinguishing between the two cases should be possible by multi-band observations with high enough cadence, since it is helpful to have a color constraint during the (sometimes short) anomalies, so we do not expect that this issue would significantly affect the current predictions of the exoplanetary yield from Roman. Nevertheless, it would be interesting to clarify and quantify this contamination with a dedicated investigation.

\acknowledgments
This work has made use of the IAC-STAR Synthetic CMD computation code. IAC-STAR is supported and maintained by the IT department of the Instituto de Astrofísica de Canarias.

The MOA project is supported by JSPS KAK-ENHI Grant Number JSPS24253004, JSPS26247023, JSPS23340064, JSPS15H00781, JP16H06287,17H02871 and 19KK0082 .

CR was supported by the ANR project COLD-WORLDS of the French \emph{Agence Nationale de la Recherche} with the reference ANR-18-CE31-0002.

\bibliography{mb06074}{}
\bibliographystyle{aasjournal}

\end{document}